\RequirePackage[2020-02-02]{latexrelease}

\documentclass[prl,floatfix,showpacs,twocolumn,preprintnumbers,amsmath,amssymb,superscriptaddress]{revtex4}

\usepackage[sort&compress]{natbib}

\usepackage{graphicx,float,datetime}
\usepackage{hyperref,wasysym}
\usepackage{epstopdf,subfigure}
\usepackage{enumitem}
\usepackage{bm}
\usepackage[utf8]{inputenc}

\usepackage{gensymb}
\usepackage{multirow}
\newcommand*\xbar[1]{%
  \hbox{%
    \vbox{%
      \hrule height 0.5pt % The actual bar
      \kern0.5ex%         % Distance between bar and symbol
      \hbox{%
        \kern-0.1em%      % Shortening on the left side
        \ensuremath{#1}%
        \kern-0.1em%      % Shortening on the right side
      }%
    }%
  }%
} 

\usepackage{array}
\newcolumntype{P}[1]{>{\centering\arraybackslash}p{#1}}
\newcolumntype{M}[1]{>{\centering\arraybackslash}m{#1}}

\def\ber{\begin{eqnarray}}
\def\eer{\end{eqnarray}}
\def\beq{\begin{equation}}
\def\eeq{\end{equation}}

\newdateformat{mydate}{\THEDAY\hspace{3pt}\monthname[\THEMONTH] \THEYEAR}

\begin{document}

\begin{center}
\title{Quantification of Captopril using Ultra High Performance Liquid Chromatography}
\date{\mydate\today}

\author{Glenn Zammit \footnote{glenn.zammit.a101215@mcast.edu.mt}}
\affiliation{Malta College of Arts, Science and Technology, Poala, Malta}

\author{Ralph Cassar \footnote{ralph.cassar@mcast.edu.mt}}
\affiliation{Malta College of Arts, Science and Technology, Poala, Malta}

\author{Mark Pace \footnote{mark.pace@mcast.edu.mt}}
\affiliation{Malta College of Arts, Science and Technology, Poala, Malta}
%\affiliation{Institute of Space Sciences and Astronomy, University of Malta, Msida, MSD 2080, Malta}

\begin{abstract}
{
\noindent
We develop an efficient, sensitive, accurate, cost effective, and greener liquid chromatography method for the determination of a quantification method concerning Captopril in pharmaceutical dosage forms by using ultra high performance liquid chromatography (UHPLC) equipment. Satisfactory resolution was able to be achieved using a Water Aquity UHPLC BEH C18 1.7 $\mu$m $\times$ 2.1 mm $\times$ 50 mm column and an isocratic mobile phase composition of Methanol, Milli-Q water and Trifluoroacetic acid in the ratio of 55:45:0.05 v/v/v respectively at a flow rate of 0.1mL/min and a wavelength detection of 220nm. The retention time for Captopril was found to be 1.744 minutes while for Captopril disulfide it was found to be 2.657 minutes. The method was validated based on the ICH and USP guidelines.
}
\end{abstract}


\maketitle

\end{center}

%------------------------Section-------------------------
\section{I. Introduction}\label{sec:intro}
%------------------------Section-------------------------
\noindent
Captopril (((2S)-1-[(2S)-2-Methyl-3-sulfanylpropanoyl] pyrrolidine-2- carboxylic acid), molecular formula C$_9$H$_{15}$NO$_3$S is a popular well-known drug which subjects in the class of drugs called angiotensin-converting enzyme (ACE) inhibitors \cite{vidt1982captopril}. Captopril is the first orally active inhibitor of ACE inhibitor in which the enzyme is responsible for the conversion of inactive angiotensin I to the potent pressor peptide angiotensin II \cite{vidt1982captopril}. ACE inhibitors are mainly used for the treatment of high blood pressure but they also contain some other medical properties such as vasculo-protective and antithrombotic activities which plays a favourable role in terms of cardiovascular morbidity \cite{bojarska2015captopril}. Captopril, has established a position in the medical treatment for hypertension and congestive heart failure \cite{amar2021captopril, sigit2020comparison}. Amongst the hypertension drugs, Captopril is a preferred drug amongst doctors, it is prescribed to patients who are chronically ill and require a long-term treatment, due to its therapeutic benefits and because of its effectiveness, low price and low toxicity \cite{vidt1982captopril, brogden1988captopril, captopril1988comparative, amar2021captopril, sigit2020comparison}.
\\
\\
Captopril disulfide (1,1$'$-[Disulfanediylbis[(2S)-2-methyl-1-oxopropane3,1- diyl]] bis[(2S)-pyrrolidine-2-carboxylic] acid), molecular formula C$_{18}$H$_{28}$N$_2$O$_6$S$_2$ is the major degradation impurity of Captopril \cite{vasconcelos2021structural}. Captopril disulfide is also referred to as impurity A \cite{vasconcelos2021structural}. The Captopril molecule starts to degrade after eight hours when in solution and also when being exposed to air by means of oxidation \cite{timmins1982factors, hillaert1999determination}. The Captopril molecule was analytically tested for stability under different stress conditions \cite{carje2019hplc}. The stability testing was performed under various pH and temperature conditions, in the presence of strong oxidizing agents and under both natural light and ultra-violet (UV) light exposure, this was done in order to distinguish it from other types of impurities that may be present in the final product. Although approved in 1981 in the USA, Captopril is still widely used today \cite{national2018livertox}.

\begin{figure}[h!]
\includegraphics[scale=0.6]{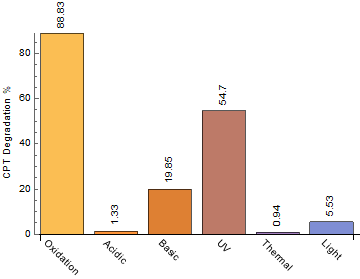}
\caption{The degradation rate of Captopril (\%) under different forced degradation conditions \cite{carje2019hplc}}
\label{CPT_deg_percentage}
\end{figure}

%\pgfplotstableread[row sep=\\,col sep=&]{
%    Method & CPT Degradation \% \\
%    Oxidation	&	88.83   \\
%    Acidic 		&	1.33	\\
%    Basic		&	19.85	\\
%    UV			&	54.7	\\
%    Thermal		&	0.94	\\
%    Light		&	5.53	
%    }\mydata

The aim of this research is to develop a cost effective, efficient and a greener liquid chromatography method for the determination of Captopril in pharmaceutical dosage forms using shorter runs and using less mobile phase than conventional liquid chromatography methods.

%--------------------Sub-Section-------------------------
\subsection{Greener instrumental analytical chemistry}
%--------------------Sub-Section-------------------------

The use of UHPLC methods in pharmaceutical analysis while reducing the time for analysis of bathces for relese to patients and customers, decreases cost of testing. Much less solvents are used. In fact this study shows that a significant reduction of 75\% of mobile phase is achieved compared to the USP method for Captopril \cite{USP}. This means that the UHPLC method proposed is greener. UHPLC methods reduce energy consumption, reduce the use of solvents, and make waste management more manageable in terms of costs and environmental impacts \cite{cielecka2013uhplc}.
\\
\\
The UHPLC method validated in this paper satisfies some of the principles of green analytical chemistry as described in \cite{galuszka201312}. In particular using UHPLC rather than a conventional HPLC method generates much less waste, reduces analysis time and the use of energy. The number of samples to test for assay in the pharmaceutical industry is determined according to guidelines such as the ICH guidelines \cite{guideline2005validation} and good manufacturing practices (GMP) \cite{nally2016good}.

%------------------------Section-------------------------
\section{II. Materials and Methods}\label{sec:materials_and_methods}
%------------------------Section-------------------------

%--------------------Sub-Section-------------------------
\subsection{Materials and reagents}
%--------------------Sub-Section-------------------------
USP grade CPT, USP grade CPT-DIS, captopril tablets, and necessary reagents.

%--------------------Sub-Section-------------------------
\subsection{Apparatus}
%--------------------Sub-Section-------------------------
Waters Acquity UHPLC and Aquity UHPLC BEH C18 1.7$\mu$m $\times$ 2.1mm $\times$ 50mm column. ACQUITY H-CLASS UHPLC.

%--------------------Sub-Section-------------------------
\subsection{Chromatographic conditions}
%--------------------Sub-Section-------------------------
Both the sample cooler temperature and the column oven temperature were set at 25$^{\circ}$C. The compounds were separated isocratically with a mobile phase composition of 550mL methanol (Carlo Erba. Gold Ultragradient grade), 450mL milli-Q water and 0.50mL trifluoroacetic acid (Fisher, 99\% HPLC Grade) having a flow rate of 0.1mL/min with an injection volume of 0.8$\mu$L. The wavelength detection for was set at 220nm.

%--------------------Sub-Section-------------------------
\subsection{Preparation of stock solutions for system suitability criteria}
%--------------------Sub-Section-------------------------
Captopril standard solution was prepared in duplicate, labelled as `standard solution 1' and `standard solution 2', by dissolving 10mg of Captopril reference standard in a 50mL volumetric flask. Dissolved and diluted up to volume using mobile phase, filtered through a 0.45$\mu$m Millipore Millex-HV Hydrophilic PVDF Filter discarding the first 6mL of filtrate. This solution contains a concentration of 0.2mg/mL of Captopril API, prepared fresh on the day of use.
\\
\\
Captopril-disulfide solution was prepared by dissolving 25mg of CPT-Dis into a 50mL volumetric flask. Dissolved and diluted up to volume using mobile phase. This solution contains a concentration of 0.5mg/mL of Captopril-disulfide. Prepared fresh on the day of use.
\\
\\
Resolution solution was prepared by dissolving 10mg of Captopril reference standard into a 50mL volumetric flask. 5mL of Captopril disulfide solution (0.5mg/mL) was added together with sufficient mobile phase. Dissolved and diluted up to volume using mobile phase, filtered through a 0.45$\mu$m Millipore Millex-HV Hydrophilic PVDF Filter discarding the first 6mL of filtrate. This solution contains a concentration of 0.2mg/mL of captopril API and 0.05mg/mL captopril disulfide, prepared fresh on the day of use.

%--------------------Sub-Section-------------------------
\subsection{Injection procedure}
%--------------------Sub-Section-------------------------
The UHPLC column was left to equilibrate with mobile phase, and blank runs were initially injected. These were followed by six repeated injections of Captopril ‘standard solution 1’ and a duplicate injection of ‘standard solution 2’. The resolution solution was injected afterwards. The six injections were used to determine the \%RSD and to calculate the \% recovery with the average area of the duplicate injection of ‘standard solution 2’. Resolution between Captopril and Captopril disulfide was determined and the tailing factor was recorded. Test solutions were bracketed with a duplicate injection of Captopril ‘standard solution 1’ and \%RSD was calculated and compared with the six repeated injection of Captopril ‘standard solution 1’ for system suitability.  

%--------------------Sub-Section-------------------------
\subsection{Sample preparation}
%--------------------Sub-Section-------------------------
The contents of 20 Captopril 100mg tablets were crushed and homogenized using a pestle and mortar. 96 mg of the Captopril crushed tablet sample were transferred to a 100mL volumetric flask. About 40mL of diluent were added, followed by sonication for 15 minutes. Dilution up to volume with diluent and filtration through a 0.45$\mu$m Millipore Millex-HV Hydrophilic PVDF Filter followed, discarding the first 6mL of filtrate. The final test solution had a concentration of 0.2mg/mL of Captopril API. 

\newpage

\begin{widetext}
\begin{center}
\begin{figure}[h!]
\includegraphics[scale=0.3]{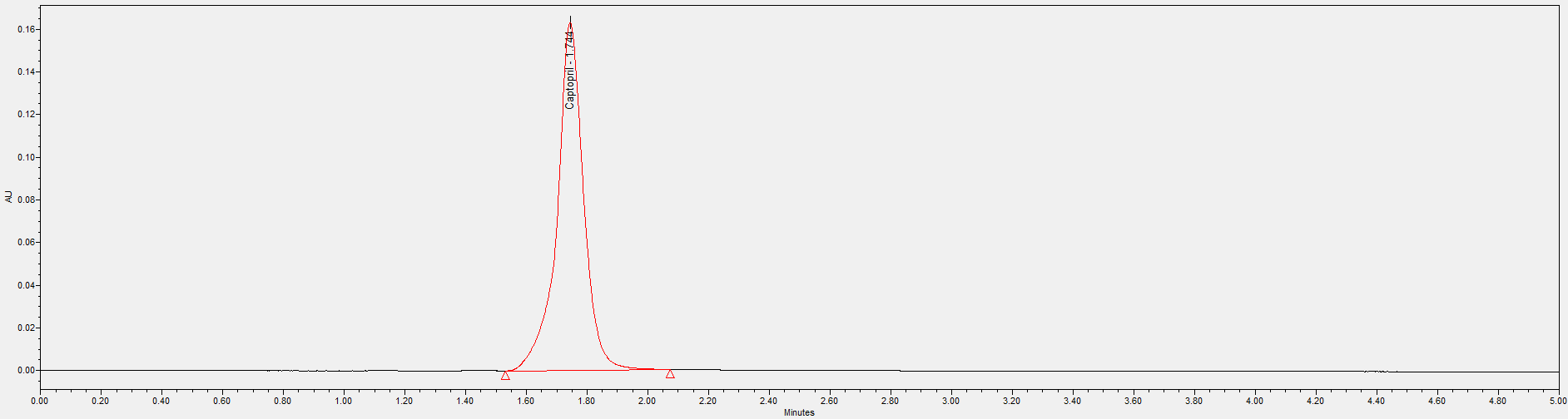}
\caption{Chromatogram of USP Captopril standard solution using UHPLC}
\label{Chromatogram_of_USP_Captopril_standard_solution}
\end{figure}

\begin{figure}[h!]
\includegraphics[scale=0.3]{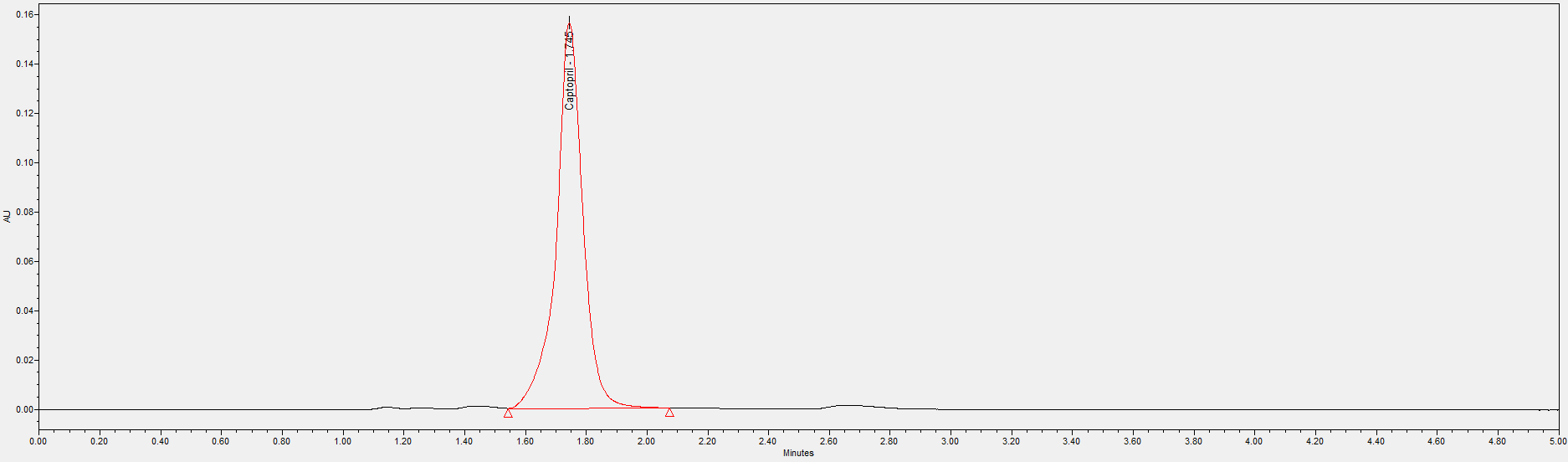}
\caption{Chromatogram of Captopril in the test solution using UHPLC}
\label{Chromatogram_of_captopril_in_the_test_solution}
\end{figure}

\begin{figure}[h!]
\includegraphics[scale=0.44]{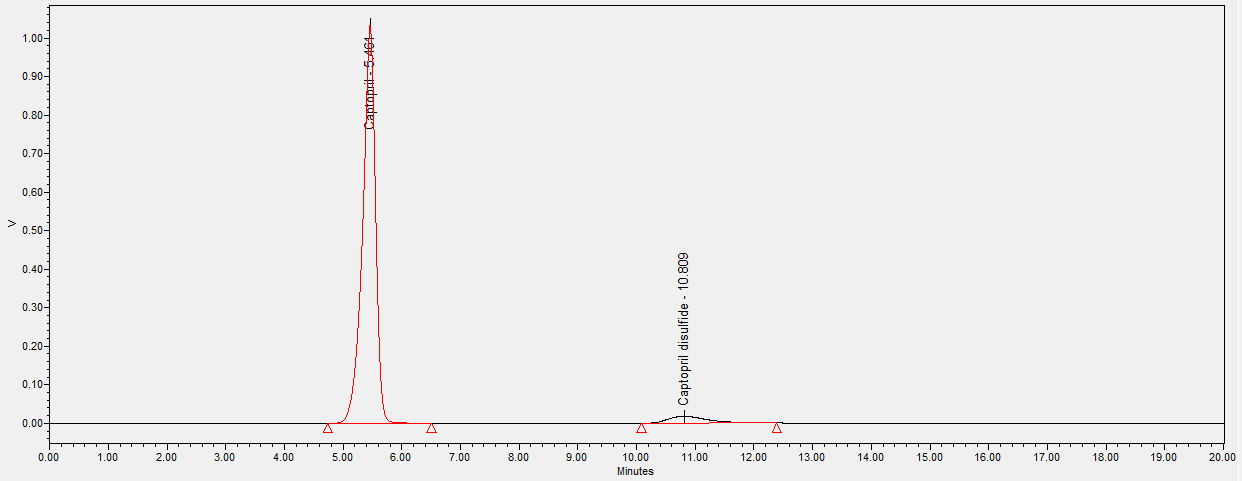}
\caption{Chromatogram of a well resolved Resolution solution showing Captopril and Captopril disulfide using UHPLC}
\label{Chromatogram_of_a_well_resolved_Resolution_solution_showing_Captopril_and_Captopril_disulfide}
\end{figure}
\end{center}
\end{widetext}

\newpage

%------------------------Section-------------------------
\section{III. Results and Discussion}
%------------------------Section-------------------------
During this study trials were conducted in order to find the optimum chromatographic conditions. The objective of this study was to develop a chromatographic method, which is more environmentally friendly compared to an HPLC method, cost effective, yet following all regulations and achieving results according to the standards set out in ICH and USP guidelines. The chromatographic method was set to achieve a peak tailing factor between 0.8 and 1.2, a retention time between 1 to 2 minutes, along with a good resolution of $>$ 1.5. A relative standard deviation of $\leq$ 2.0\% and a similarity factor between ‘standard 1’ and ‘standard 2’ injections between 98.0 and 102.0\%. The peaks for Captopril molecule and its dimer, Captopril disulfide were well defined, resolved and free from tailing. The elution order was Captopril at 1.744 minutes and Captopril disulfide at 2.657 minutes for the UHPLC method developed, whilst for the USP HPLC method the elution order is Captopril at 5.461 minutes and Captopril disulfide at 10.809 minutes.

\begin{widetext}
\begin{table}[h!]
  \centering
\begin{tabular}{ |p{7cm}|M{3.5cm}|M{3.5cm}|  }
 \hline
 \multicolumn{3}{|c|}{System suitability parameters} \\ \hline
 Parameter 							& 	Limit 			& 	Result 			\\ \hline
 Resolution of CPT with CPT-DIS		&	$>$1.5			&	4.3				\\ \hline
 \%RSD on 6 injections of STD 1		&	$\leq$ 2.0\%	&	0.5\%			\\ \hline
 Tailing factor						&	0.8 - 1.2		&	0.9				\\ \hline
 Bracketing check					&	98.0 - 102.0\%	&	99.5 - 99.8\%	\\ \hline
 Similarity factor STD 1 and STD 2	&	98.0 - 102.0\%	&	100.9\%			\\ \hline

\end{tabular}
\caption{System suitability parameters}
\end{table}
\end{widetext}

Running a single batch for analyses using the HPLC and UHPLC methods consist of 17 runs which includes; 2 blank injections at the beginning, 6 injections of ‘standard solution 1’, 2 injections of ‘standard solution 2’ followed by a single resolution injection. Then another 2 ‘standard solution 1’ injections, batch sample in duplicate and another 2 ‘standard solution 1’ injections to close off the bracketing of standards. 
\\
\\
HPLC method: Running the above sample set, a mobile phase usage of 340mL for the analysis of one batch.
\\
\\
Developed UHPLC method: Running the same sample set using the UHPLC method parameters, a mobile phase usage of 8.5mL for the analysis of one batch.
\\
\\
On another note, the Captopril USP monograph declares that Captopril solutions have a stability of 8 hours, meaning that when using the conventional USP HPLC method only a total of 24 injections i.e., a maximum of 4 batch runs can be conducted when excluding the 2 blank injections at the beginning of the run. Whilst with the newly developed UHPLC method it is possible to run a total of 96 injections i.e., a maximum of 31 batches in a single 8-hour sample set, excluding the 2 blank injections at the beginning of the run. This clearly shows that the new greener UHPLC method is far more efficient and requires a much less solvent to carry out the quantification analysis.  

\newpage

\begin{widetext}
\begin{table}[h!]
  \centering
\begin{tabular}{ |m{3.5cm}|M{3.5cm}|M{3.5cm}|p{3.5cm}|  }
 \hline
 \multicolumn{4}{|c|}{Comparing HPLC vs UHPLC method when conducting one batch of quantification analysis} \\ \hline
 Parameter 				& 	HPLC method	 & 	UHPLC method	&	Results in \%	\\ \hline
 Run time (minutes)		&	20			 &	5				&	UHPLC is 75\% more efficient when compared to HPLC method	\\ \hline
 Mobile phase used (mL)	&	340			 &	8.5				&	97.5\% less mobile phase was used for the UHPLC analysis	\\ \hline
 
\end{tabular}
\caption{HPLC and UHPLC methods comparison}
%\end{table}
%
%\begin{table}[h!]
%  \centering
\begin{tabular}{ |M{7cm}|M{3.6cm}|M{3.6cm}|  }
 \hline
 \multicolumn{3}{|c|}{Batch runs that can be conducted within 8 hours for stability tests} \\ \hline
 HPLC method	 		& 	UHPLC method		& 	Percentage 		\\ \hline
 4						&	31					&	775\% more batches for the UHPLC method \\ \hline

\end{tabular}
\caption{Batch run times for HPLC and UHPLC methods}
\end{table}
\end{widetext}

%------------------------Section-------------------------
\section{IV. Method Validation Results}
%------------------------Section-------------------------
%--------------------Sub-Section-------------------------
\subsection{Specificity}
%--------------------Sub-Section-------------------------
The specificity tests, for any interference found in the excipients (placebo), mobile phase, blank, and degradation impurity with the main analyte Captopril. As chromatograms were overlayed together the test is said to comply because there was no interference from excipients and degradant impurity Captopril disulfide with the main analyte of interest, Captopril, retention time 1.744 minutes.

\begin{widetext}
\begin{center}
\begin{figure}[h!]
\includegraphics[scale=0.45]{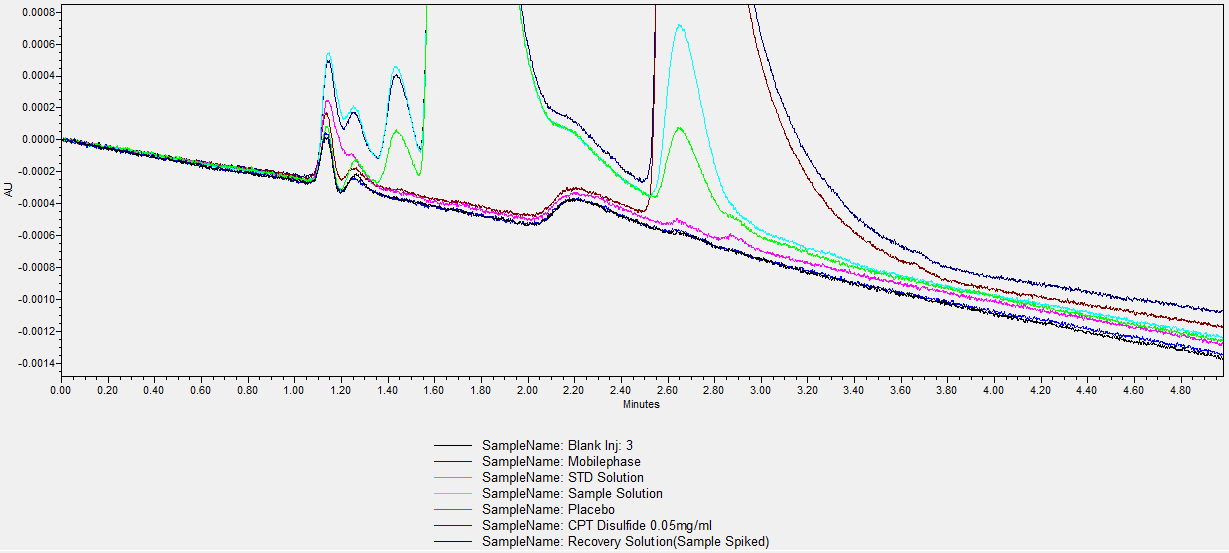}
\caption{An overlay chromatogram showing the specificity test}
\end{figure}
\end{center}
\end{widetext}

%--------------------Sub-Section-------------------------
\subsection{Linearity and range}
%--------------------Sub-Section-------------------------

The linearity was determined by injecting a series of 5 standard solutions whose concentrations were at 0.05mg/mL, 0.10mg/mL, 0.15mg/mL, 0.2mg/mL (Target concentration, 100\%), and 0.25mg/mL. As shown in Fig. (\ref{Graph_of_area_against_concentration}) a plot of area of response (UNITS) vs linearity concentration levels in mg/mL was plotted. The correlation coefficient (R), range and the residual sum of squares graph were determined. All values comply with the ICH and USP guidelines. The values of the correlation coefficient were close to unity which indicates a good linearity and shows that the developed UHPLC method has a high sensitivity.

\begin{widetext}
\vspace{1cm}

\begin{center}
\begin{figure}[h!]
\includegraphics[scale=0.55]{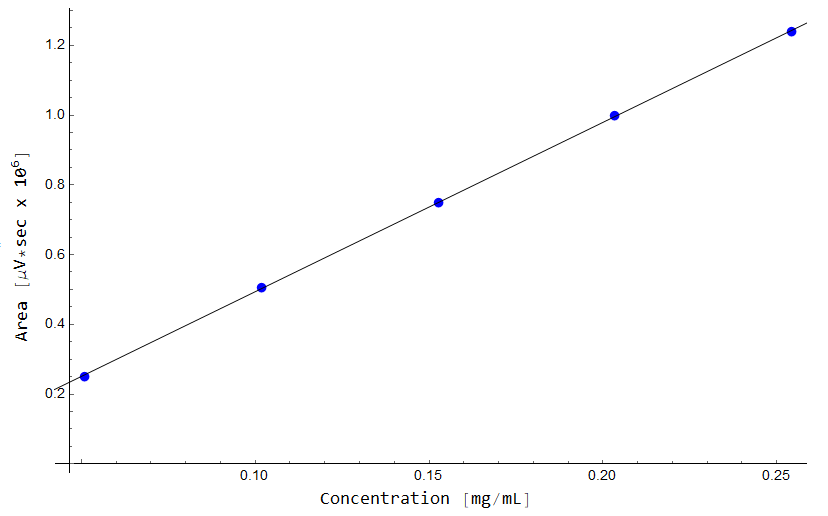}
\caption{Graph of Area [$\mu V*$ sec $\times$ 10$^6$] against Concentration [mg/mL]}
\label{Graph_of_area_against_concentration}
\end{figure}

\vspace{1.5cm}

\begin{figure}[h!]
\includegraphics[scale=0.65]{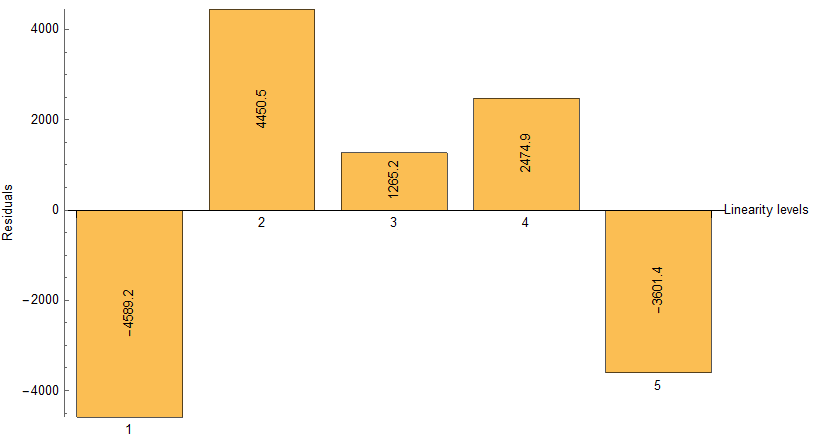}
\caption{Residuals Scattered Plot}
\label{Residuals_Scattered_Plot}
\end{figure}
\end{center}

%--------------------Sub-Section-------------------------
\subsection{Summary of results}
%--------------------Sub-Section-------------------------

\begin{table}[h!]
\begin{center}
\begin{tabular}{|c|c|c|ll}
\cline{1-3}
Parameter                                                       & ICH Guidelines Limit              & Result                                   &  &  \\ \cline{1-3}
\multicolumn{1}{|l|}{Correlation coefficient (R)}               & \textgreater 0.995                & 0.99995                                  &  &  \\ \cline{1-3}
\multicolumn{2}{|c|}{The   residual sum of squares should be randomly distributed and show no bias} & Complies,   distributed and show no bias &  &  \\ \cline{1-3}
\multicolumn{2}{|c|}{Residual   Sum of Squares (RSS)}                                               & 61563649.90                              &  &  \\ \cline{1-3}
\multicolumn{2}{|c|}{Range   (Upper and lower intervals)}                                           & 25 to 125\%                              &  &  \\ \cline{1-3}
\end{tabular}
\caption{Results gained compared to ICH guidelines}
\end{center}
\end{table}
\end{widetext}

\subsection{Accuracy}

%Accuracy is defined as the closeness of agreement between a test result and the accepted reference value within a combination of random and systematic errors. 
The accuracy of a method was determined by performing a recovery experiment. Concentration levels and preparations were done according to the ICH and USP guidelines. For 70\% and 130\% concentration levels, 3 preparations were done, and for the 100\% concentration levels 6 preparations were prepared and injected. The percentage recovery i.e. the actual concentration divided by the theoretical concentration multiplied by a 100 was calculated. The percent recovery of each individual should be $>$ 98.0\%. The \%RSD was determined for each concentration level where the limit is $<$ 2.0\%.
\\
\\
None of the individual samples were less than 98.0\% and the \%RSD was found to be less than 2.0\% for all concentration levels. This indicates a good repeatability and reliability of the proposed UHPLC method.

\begin{widetext}
\begin{center}
\begin{table}[h!]
\begin{tabular}{|l|l|l|l|}
\hline
Accuracy Concentration Levels & \%Recovery & \%Mean                                       & \%RSD                \\ \hline
Accuracy 70\% S1              & 101.11     & \multicolumn{1}{c|}{\multirow{3}{*}{100.29}} & \multirow{3}{*}{1.1} \\ \cline{1-2}
Accuracy 70\% S2              & 99.08      & \multicolumn{1}{c|}{}                        &                      \\ \cline{1-2}
Accuracy 70\% S3              & 100.69     & \multicolumn{1}{c|}{}                        &                      \\ \hline
Accuracy 100\% S1             & 100.35     & \multirow{6}{*}{99.30}                       & \multirow{6}{*}{0.8} \\ \cline{1-2}
Accuracy 100\% S2             & 98.42      &                                              &                      \\ \cline{1-2}
Accuracy 100\% S3             & 99.18      &                                              &                      \\ \cline{1-2}
Accuracy 100\% S4             & 99.35      &                                              &                      \\ \cline{1-2}
Accuracy 100\% S5             & 98.36      &                                              &                      \\ \cline{1-2}
Accuracy 100\% S6             & 100.13     &                                              &                      \\ \hline
Accuracy 130\% S1             & 101.36     & \multirow{3}{*}{99.50}                       & \multirow{3}{*}{1.7} \\ \cline{1-2}
Accuracy 130\% S2             & 99.13      &                                              &                      \\ \cline{1-2}
Accuracy 130\% S3             & 98.02      &                                              &                      \\ \hline
\end{tabular}

\caption{Accuracy concentration levels}
\end{table}
\end{center}
\end{widetext}

\subsection{Precision and intermediate precision}

%Precision of an analytical method demonstrate the closeness of agreement between a series of measurements obtained from multiple sampling of the same homogeneous sample.
The precision testing was performed by injecting six determinations of batch sample at a concentration level of 100\% (0.2mg/mL). It was imperative that the limits for the precision for individual values fall between 90.0 to 110.0\% of the labelled amount and the \%RSD on the six injections should be $<$ 2.0\%. 
%Results, complies indicating a good repeatability and reliability of results. 

%\begin{widetext}
%\begin{center}
%\begin{table}[h!]
%\begin{tabular}{|l|l|l|l}
%\cline{1-3}
%Precision    & \%Assay & \%RSD on 6 samples   &  \\ \cline{1-3}
%Precision S1 & 96.91   & \multirow{6}{*}{0.7} &  \\ \cline{1-2}
%Precision S2 & 96.96   &                      &  \\ \cline{1-2}
%Precision S3 & 97.98   &                      &  \\ \cline{1-2}
%Precision S4 & 98.56   &                      &  \\ \cline{1-2}
%Precision S5 & 97.36   &                      &  \\ \cline{1-2}
%Precision S6 & 97.02   &                      &  \\ \cline{1-3}
%\end{tabular}
%
%\end{table}
%\end{center}
%\end{widetext}

The intermediate precision tests were conducted by a different analyst, on different equipment and on a different day.
%The purpose of intermediate precision or ruggedness tests is to determine whether within laboratory variations and conditions, such as the analysis being performed by a different analyst, on a different day, using different equipment, and a new UHPLC column, new mobile phases, diluents, and sample preparation will affect precision. 
\\
\\
These results were gathered from 6 injections of precision and 6 of intermediate combined together. The limits for intermediate precision is that of $<$ 2.0\%.
%The limits for the intermediate precision include, that the individual assay value fall between 90.0 to 110.0\% of the labelled amount and the \%RSD on the 12 injections (including the 6 injections of the precision test) should fall < 2.0\%. The intermediate precision was carried out by another user by the use of a Waters ACQUITY H-CLASS UHPLC, and a new UHPLC column, AQUITY UHPLC BEH C18 1.7$\mu$m x 2.1mm x 50mm. 

\begin{widetext}
\begin{center}
% \usepackage{multirow}
\begin{table}[h!]
\begin{tabular}{|l|c|c|c|}
\hline
\textbf{Precision}        & \multicolumn{1}{l|}{\textbf{\%Assay}} & \multicolumn{1}{l|}{\textbf{\%RSD on 6 samples}} & \multicolumn{1}{l|}{\textbf{\%RSD on 12 samples}} \\ \hline
Precision S1              & 96.91                                 & \multirow{6}{*}{0.7}                             & \multirow{12}{*}{0.8}                             \\ \cline{1-2}
Precision S2              & 96.96                                 &                                                  &                                                   \\ \cline{1-2}
Precision S3              & 97.98                                 &                                                  &                                                   \\ \cline{1-2}
Precision S4              & 98.56                                 &                                                  &                                                   \\ \cline{1-2}
Precision S5              & 97.36                                 &                                                  &                                                   \\ \cline{1-2}
Precision S6              & 97.02                                 &                                                  &                                                   \\ \cline{1-3}
Intermediate Precision S1 & 98.61                                 & \multirow{6}{*}{0.7}                             &                                                   \\ \cline{1-2}
Intermediate Precision S2 & 97.86                                 &                                                  &                                                   \\ \cline{1-2}
Intermediate Precision S3 & 99.16                                 &                                                  &                                                   \\ \cline{1-2}
Intermediate Precision S4 & 98.24                                 &                                                  &                                                   \\ \cline{1-2}
Intermediate Precision S5 & 98.33                                 &                                                  &                                                   \\ \cline{1-2}
Intermediate Precision S6 & 97.27                                 &                                                  &                                                   \\ \hline
\end{tabular}
\caption{Precision of results}
\end{table}

\end{center}
\end{widetext}

\section{V. Conclusion}

In this study we have developed a new method for quantifying captopril using UHPLC. The study shows that this is may be a viable way forward for the pharmaceutical industry, given the increasing emphasis on  environmental responsibility and efficient and cost-effective use of resources. This validation study of the new method was carried out according the ICH and USP guidelines.

\noindent

\section{VI. Acknowledgements}
\noindent
We would like to thank Combino Pharm Malta Ltd. for their invaluable contributions to this study.

%LEAVE HERE FOR CORRECTION!! JUST COPY BBL FILE NOW
%\bibliographystyle{IEEEtran}
%\bibliography{Bibliography}

\end{document}